\documentclass[12pt]{emulateapj}
\usepackage{amsmath,amssymb}

\newcommand{\vct}[1]{\mbox{\boldmath{$#1$}}}
\newcommand{\Flin}{F_{\rm lin}}
\newcommand{\M}{\mathcal{M}}
\newcommand{\A}{\mathcal{A}}
\newcommand{\B}{\mathcal{B}}
\newcommand{\R}{\mathcal{R}}
\newcommand{\I}{\mathcal{I}}
\newcommand\Msun{{\;\rm\,M_\odot}}
\newcommand\kms{{\;\rm km\,s^{-1}}}

\newcommand{\rs}{r_s}
\newcommand{\Rp}{R_p}

\newcommand{\torb}{t_{\rm orb}}
\newcommand{\tcr}{t_{\rm cross}}
\newcommand{\cs}{a_{0}}
\newcommand{\rmin}{r_{\rm min}}
\newcommand{\MP}{M_{\scriptscriptstyle P}}
\newcommand{\PhiP}{\Phi_{\scriptscriptstyle P}}

\newcommand{\etaA}{\eta_{\scriptscriptstyle \mathcal{A}}}
\newcommand{\etaB}{\eta_{\scriptscriptstyle \mathcal{B}}}
\newcommand\simlt{\lower2pt
   \hbox{$\;\buildrel{\scriptstyle <}\over{\scriptstyle\sim} \; $}}
\newcommand\simgt{\lower2pt
   \hbox{$\;\buildrel{\scriptstyle >}\over{\scriptstyle\sim} \; $}}
\newcommand{\taud}{\tau_{\rm{decay}}}

\shorttitle{Nonlinear Dynamical Friction}
\shortauthors{W.-T.~Kim}

\begin{document}
\title{Nonlinear Dynamical Friction of a Circular-Orbit Perturber
in a Gaseous Medium}
\author{Woong-Tae Kim$^{1,2}$}
\affil{$^1$Center for the Exploration of the Origin of the Universe
(CEOU), Astronomy Program, Department of Physics \& Astronomy, Seoul
National University, Shillim-Dong, Kwanak-Gu, Seoul 151-742,
Republic of Korea \\ $^2$Astronomy Program, FPRD, Department of
Physics and Astronomy, Seoul National University, Shillim-Dong,
Kwanak-Gu, Seoul 151-742, Republic of Korea}
\email{wkim@astro.snu.ac.kr}

\begin{abstract}
We use three-dimensional hydrodynamic simulations to investigate the
nonlinear gravitational responses of gas to, and the resulting drag
forces on, very massive perturbers moving on circular orbits. This
work extends our previous studies that explored the cases of
low-mass perturbers on circular orbits and massive perturbers on
straight-line trajectories.  The background medium is assumed to be
non-rotating, adiabatic with index $5/3$, and uniform with density
$\rho_0$ and sound speed $\cs$. We model the gravitating perturber
using a Plummer sphere with mass $M_p$ and softening radius $\rs$ in
a uniform circular motion at speed $V_p$ and orbital radius $\Rp$,
and run various models with differing $\R\equiv \rs/\Rp$, $\M\equiv
V_p/\cs$, and $\B\equiv GM_p/(\cs^2\Rp)$. A quasi-steady density
wake of a supersonic model consists of a hydrostatic envelope
surrounding the perturber, an upstream bow shock, and a trailing
low-density region. The continuous change in the direction of the
perturber motion makes the detached shock distance reduced compared
to the linear-trajectory cases, while the orbit-averaged gravity of
the perturber gathers the gas toward the center of the orbit,
modifying the background preshock density to $\rho_1 \approx
(1+0.46\B^{1.1})\rho_0$ depending weakly on $\M$. For sufficiently
massive perturbers, the presence of a hydrostatic envelope makes the
drag force smaller than the prediction of the linear perturbation
theory, resulting in $F=4\pi\rho_1(GM_p)^2/V_p^2 \times
(0.7\etaB^{-1})$ for $\etaB\equiv \B/(\M^2-1) > 0.1$; the drag force
for low-mass perturbers with $\etaB<0.1$ agrees well with the linear
prediction. The nonlinear drag force becomes independent of $\R$ as
long as $\R<\etaB/2$, which places an upper limit on the perturber
size for accurate evaluation of the drag force in numerical
simulations.
\end{abstract}


\section{Introduction}\label{sec:intro}

Dynamical friction (DF) arising from the gravitational interaction
of a massive body with a background medium is considered to be one
of the most powerful mechanisms responsible for the orbital decay of
astronomical objects (e.g., \citealt{bin08} and references therein).
DF occurs not only in a collisionless background in which
long-range, two-body interactions transfer momentum from an object
in motion to the field particles, but also in a continuous gaseous
medium where a perturber loses its momentum due to the gravitational
drag force exerted by its own induced wake.  The DF in a gaseous
background is likely to play an important role in the orbital decay
of companions in common-envelope binaries, supermassive black holes
(SMBHs) at galaxy centers, massive galaxies in galaxy clusters, etc.
For instance, in a binary system, DF causes a low-mass companion
engulfed by an extended primary to spiral in toward the core of the
primary, which in turn spins up the primary envelope, greatly
affecting the subsequent evolution of the system (e.g.,
\citealt{taa00,edg04,nor06,max09} and references therein). Also,
numerical simulations show that the DF due to a gaseous medium
expedites the growth of SMBHs by mergers in colliding galaxies
(e.g., \citealt{mil03,esc04,esc05,dot06,may07,col09}), potentially explaining the ubiquity of SMBHs at galactic nuclei (e.g.,
\citealt{beg80,men01,fer05}).

Using a time-dependent linear perturbation theory,
\citet{ost99} showed that the drag force on a
perturber with mass $M_p$ moving at speed $V_p$ on a rectilinear
trajectory through a uniform gaseous  medium with density $\rho_0$ and
sound speed $\cs$ is given by
\begin{equation}\label{eq:f_pt_linear}
  \Flin = \frac{4\pi\rho_0(GM_p)^2}{V_p^2}
  \left\{\begin{array}{ll}
     \frac{1}{2}\ln\left(\frac{1+\M}{1-\M}\right)-\M, & \M<1,\\
     \frac{1}{2}\ln\left(1-\frac{1}{\M^2}\right)
               +\ln\left(\frac{V_p t}{\rmin}\right), & \M>1,
  \end{array}\right.
\end{equation}
where $\M\equiv V_p/\cs$ is the Mach number, and $\rmin$ is the
characteristic size of the perturber. The key features of equation
(\ref{eq:f_pt_linear}) are that (i) the gaseous DF force becomes
identical to the classical formula of \citet{cha43} for the
collisionless drag when $\M\gg1$, (ii) the gaseous DF force is more
efficient than the collisionless drag for $\M\sim1$, and (iii) the
gaseous DF force is nonvanishing even for subsonic perturbers with
$\M<1$ (see also \citealt{jus90}). The last point improves the
previous notion that the gaseous drag is absent in subsonic cases
due to the front-back symmetry in the steady-state wakes (e..g,
\citealt{dok64,rud71,rep80}). Several recent studies showed that
equation (\ref{eq:f_pt_linear}) can be applicable, with some
modifications, to more general cases such as in a radially
stratified medium \citep{san01}, for circular-orbit perturbers
\citep{kk07,kim08}, for perturbers with relativistic speed
\citep{bar07} or in accelerating motion \citep{nam10}, etc. In
particular, \citet[hereafter KK07]{kk07} used a semi-analytic method
to show that equation (\ref{eq:f_pt_linear}) is a good approximation
to the DF force on circular-orbit perturbers provided $V_p t=2R_p$,
where $R_p$ is the orbital radius.

While all the theoretical studies mentioned above consider low-mass
perturbers and find various astrophysical applications (e.g.,
\citealt{nar00,elz04,kim05,kim07,con08,vil09}), there are some
situations such as in orbital decay of SMBHs or companions in
common-envelope binaries, where perturbers have so large masses that
the induced density wakes are in the nonlinear regime. Using
hydrodynamic simulations, \citet[hereafter KK09]{kk09} extended the
work of \citet{ost99} to study nonlinear DF force for a very massive
perturber on a straight-line trajectory. By modeling a perturber
using a Plummer sphere with softening radius $\rs$, KK09 found that
a nonlinear supersonic wake is characterized by a detached bow shock
and a hydrostatic envelope near the perturber. The resulting drag
force depends solely on the dimensionless parameter $\etaA$ defined
as
\begin{equation}\label{etaA}
  \etaA \equiv \frac{\A}{\M^2-1}, {\rm\;\;\;with \;\;}
  \A \equiv \frac{GM_p}{\cs^2\rs},
\end{equation}
and is given by
\begin{equation}\label{eq:KK09}
  \frac{F}{\Flin} = \left(\frac{\etaA}{2}\right)^{-0.45},
  {\rm \;\;\;for\;\;\etaA>2},
\end{equation}
while $F/\Flin \approx 1$ for $\etaA<0.7$. The reduction of the
nonlinear DF force compared to the linear estimate is due to the
presence of a hydrostatic envelope that makes the density
distribution spherically symmetric near the perturber.  This clearly
demonstrates that the nonlinear effect can be significant if a
perturber is very massive.

Since astronomical objects usually follow curvilinear rather than
straight-line trajectories, it is interesting to see whether
equation (\ref{eq:KK09}) remains valid for perturbers on circular
orbits. For this purpose, we in this paper take one step further
from KK09 to study the nonlinear gravitational responses of gas to,
and the associated drag force on, a very massive perturber moving on
a circular orbit. This work also extends the linear perturbation
analyses of KK07 by considering the nonlinear effect on the density
wakes. The intensity of gravitational influence of a perturber to
the background gas can be measured by the Bondi radius $r_{\rm
B}=GM_p/\cs^2$. For perturbers on straight-line trajectories, the
softening radius (or the perturber size) is the lone length scale
with which the Bondi radius can be compared. The drag force,
correspondingly, depends on $\rs$ and $M_p$ only through the
dimensionless parameter $\A$: increasing $\rs$ is equivalent to
decreasing $M_p$. Equation (\ref{eq:KK09}) then predicts that the
drag force on a highly nonlinear object with $\A \gg 1$ would be
negligible, which results simply from the fact that $\rs$ is
inseparable from $M_p$. On the other hand, circular orbits with
orbital radius $R_p$ naturally introduce an additional length scale,
so that the dependence of the drag force on $\rs$ can be explored
independently of the perturber mass.

While a perturber in real astronomical situations is likely to move
through a background medium that is rotating and/or stratified, we
in this work idealize the gaseous medium as being initially static
and uniform. This certainly introduces a few important caveats that
should be noted from the outset. If the gaseous medium is supported
primarily by rotation, as in protoplanetary disks, the perturber
launches spiral waves at Lindblad resonances that propagate away
from it (e.g., \citealt{war97,tan02,cha09,lub10} and references
therein), which is quite different from quasi-steady density wakes
with long trailing tails produced in a non-rotating medium (e.g.,
KK07). If the medium is instead supported by thermal pressure, it
should have a stratified density in the radial direction, as in
intracluster media or common-envelope binaries. In this case, the
gradient in the background density profile can be ignored if the
Bondi radius is much smaller than \emph{both} the pressure scale
length \emph{and} the orbital radius, that is, if $\M \ll
(M(R_p)/M_p)^{1/4}$ and $\B \ll1$, where $M(R_p)$ is the dynamical
mass enclosed within $R_p$ and $\B$ is the dimensionless perturber
mass defined in \S\ref{sec:code} below. Although the first condition
is readily met, for example, in common-envelope binaries with
planets or brown dwarfs as low-mass companions (e.g.,
\citealt{nor06}), the second condition is not well satisfied
especially for very massive perturbers with $\B\simgt 1$ considered
in this paper. In the latter case, the wakes and the associated DF
forces are likely to be affected by the density gradient of the
background medium (e.g., see \citealt{jus05} for collisionless
cases). Neglecting the density stratification in the background
medium also suppresses gas buoyancy, precluding the potential
effects of convective motions and gravity modes in heating the
medium by resonant excitations (e.g., \citealt{bal90,luf95,kim07})
and other processes.

In addition to the above assumptions, we ignore the orbital motion
of the background gas with respect to the center of mass of the
whole system, which can be a valid approximation only when $\B/\M^2
\ll 1$. If this condition is not satisfied, the neglect of the
centrifugal and Coriolis forces arising from the orbital motion of
the background may affect the density wakes and the drag forces
(e.g., \citealt{ada89,ost92}). We also treat the gas using an
adiabatic equation of state, which implicitly assumes that the
orbital energy of the perturber is converted to heat as it spirals
inward, and thus is valid only if radiative loss is negligible. By
ignoring self-gravity, we do not consider any back reaction of the
gas on the perturber.

Given these limitations and constraints, we by no means attempt to
apply the results of this work to real astronomical systems.
Nevertheless, the idealized models considered in this paper help to
isolate the effect of the perturber size or orbital radius on the
nonlinear DF force. The results of this work will be particularly
useful to justify large perturber sizes employed in recent
hydrodynamic simulations, such as for SMBHs at galactic nuclei
(e.g., \citealt{esc04,esc05,dot06,dot07,may07,cua09}) and companions
in common-envelope binaries (e.g., \citealt{ruf93,san98,ric08}),
etc. These simulations usually treat the perturber using a softened
point mass, with its size inevitably limited by numerical
resolution. For instance, $N$-body/SPH simulations for the orbital
decay of SMBHs take quite large values (up to a few pc) for $\rs$
(e.g., \citealt{esc04,may07}), although the realistic values are
probably of order of the Schwartzschild radius ($\sim1$ AU for
$M_p=2\times10^7\Msun$). Grid-based simulations of the
common-envelope phase of binaries also represent a companion using a
point mass with size set by the grid spacing, which is larger than
the actual companion size by one or two orders of magnitude (e.g.,
\citealt{san98,ric08}). If the drag force on circular-orbit
perturbers depends on $\rs$ similarly to the linear-trajectory
cases, simulations with such large $\rs$ would significantly
underestimate the real decay time since the induced density wake
would erroneously be in the linear regime. If the drag force instead
turns out insensitive to $\rs$ in circular-orbit cases, a large
value of $\rs$ would be reasonable as long as it does not change the
drag force much.

The paper is organized as follows: In \S\ref{sec:code}, we describe
numerical methods and models we adopt. In \S\ref{sec:linear}, we run
models for low-mass perturbers and compare the resulting drag forces
with the analytical predictions to find a proper relationship
between $\rs$ and $\rmin$. In \S\ref{sec:nonl}, we present evolution
and quasi-steady distributions of fully nonlinear density wakes and
the associated drag forces on massive, circular-orbit perturbers.
Finally in \S\ref{sec:diss}, we summarize our findings and discuss
their implications on the proper choice of the perturber size.

\begin{figure*}
  \epsscale{1}
  \plotone{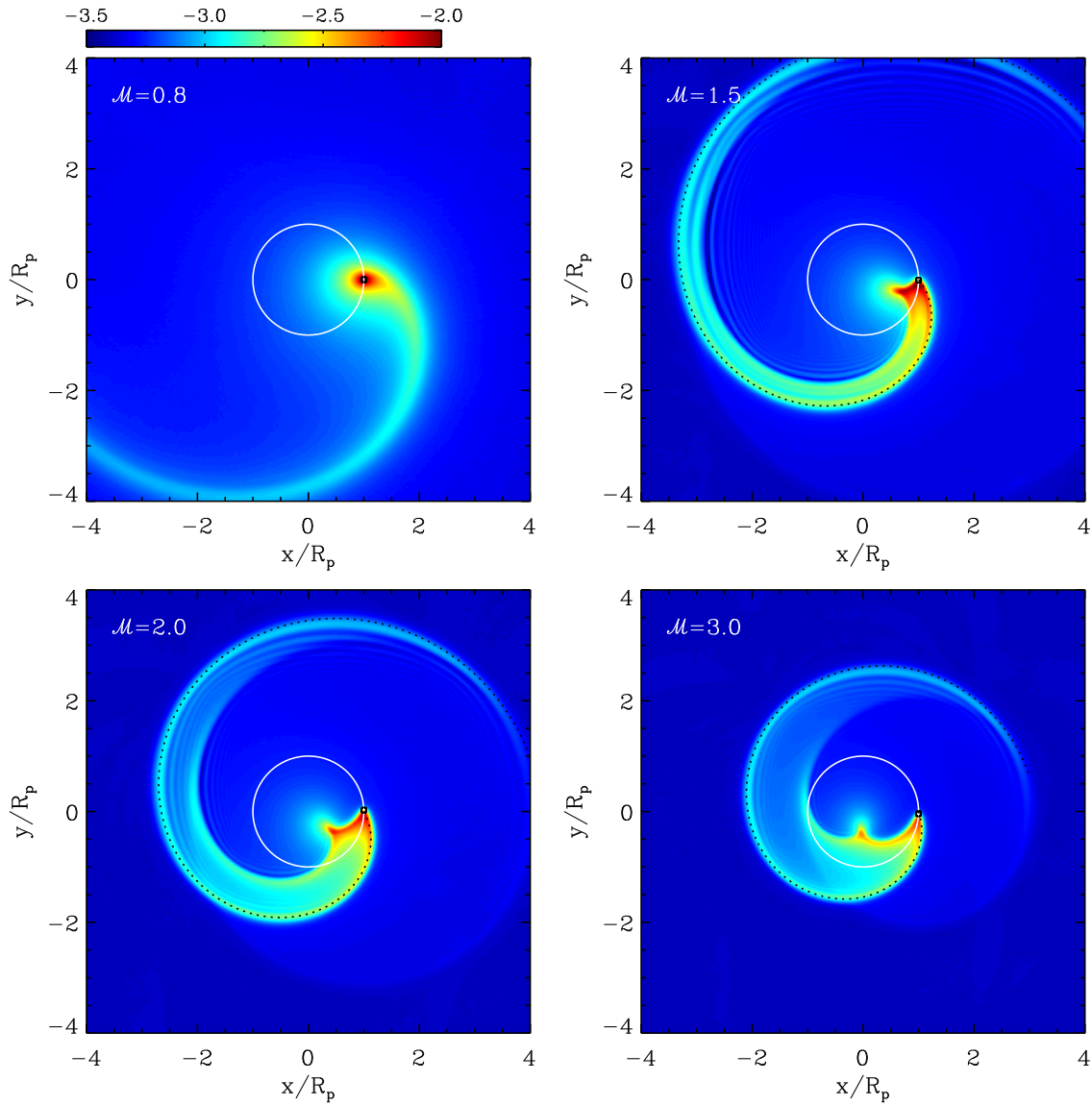}
  \caption{\label{fig:lin_wake}
  Steady-state distributions of the perturbed density
  in logarithmic color scale on the $x$--$y$ plane at $t/\torb=1$
  for low-mass perturbers
  with $\R=0.05$ and $\B=5\times 10^{-4}$ ($\A=0.01)$.
  Colorbar labels $\log (\rho/\rho_0-1)$.
  The white circle in each panel denotes the orbit, while the small black
  circle near $(x,y)=(R_p,0)$ marks the softening radius of the perturber. The spiral dotted lines
  for the supersonic cases draw the outer edges of the tails
  from the linear theory, eq.\ (B2) of KK07, which are in good agreement
  with the numerical results.
} \vspace{0.5cm}
\end{figure*}

\section{Numerical Method} \label{sec:code}

We consider a massive perturber moving on a circular orbit through a
gaseous medium, and study the gravitational response of the gas to
the perturber and the resulting drag force. Similar work for
low-mass perturbers on circular orbits and massive perturbers on
straight-line trajectories was presented in KK07 and KK09,
respectively. Since the DF timescale is usually much longer than the
orbital time, the idealized circular orbit is a reasonable
approximation to real curvilinear orbits. We assume that the
background medium is unmagnetized, non-self-gravitating, and
initially static and homogeneous with density $\rho_0$ and adiabatic
sound speed $\cs$. We adopt an adiabatic equation of state with
index $\gamma=5/3$. The perturber with mass $M_p$ moves with a
constant angular velocity $\Omega_p$ along a circle with radius
$\Rp$ on the $z=0$ plane; the corresponding linear velocity is
$V_p=\Rp\Omega_p$. We represent the perturber using a Plummer
potential
\begin{equation}\label{eq:Plummer}
\PhiP(\vct{x},t)=-\frac{GM_p}
{(\vert\vct{x}-\vct{x_p}(t)\vert^2+\rs^2)^{1/2}},
\;\;\;{\rm for}
\;t\geq0,
\end{equation}
where $\rs$ is the softening radius and
$\vct{x_p}(t)=\Rp(\cos\Omega_p t, \sin\Omega_p t, 0)$
is the perturber location at time $t$ in the Cartesian
coordinates $(x,y,z)$.

We take $\Rp$, $\cs$, and $\tcr=\Rp/\cs$ as the units of length, velocity,
and time, respectively, in our simulations.  Then, our models are
completely characterized by three dimensionless parameters:
$\R\equiv\rs/\Rp$, $\M\equiv V_p/\cs$, and
\begin{equation}\label{eq:B}
\B\equiv\frac{GM_p}{\cs^2\Rp}=\R\A,
\end{equation}
which is the ratio of the Bondi radius to the orbital radius. Note
that $\R$ is a new parameter introduced by the circular orbit, which
is absent for perturbers on straight-line trajectories. For future
purposes, we define a dimensionless parameter
\begin{equation}
\etaB \equiv \frac{\B}{\M^2-1},
\end{equation}
similarly to $\etaA$ in equation (\ref{etaA}). To explore the
parametric dependence of the drag force on $\R$, $\M$, and $\B$ (or
equivalently $\A$), we run a total of 65 three-dimensional
simulations with $\R$ varying in the range of $0.025$--$0.4$, $\M$
in $0.5$--4, and $B$ from $5\times 10^{-4}$ to $4$ ($\A$ from 0.01
to 80). Our fiducial model has $\R=0.05$, $\M=2$, and $\B=0.5$. We
use the orbital time $\torb=2\pi R_p/V_p = 2\pi\tcr/\M$ as the time
unit of our presentation.

\begin{figure*}
  \epsscale{1}
  \plotone{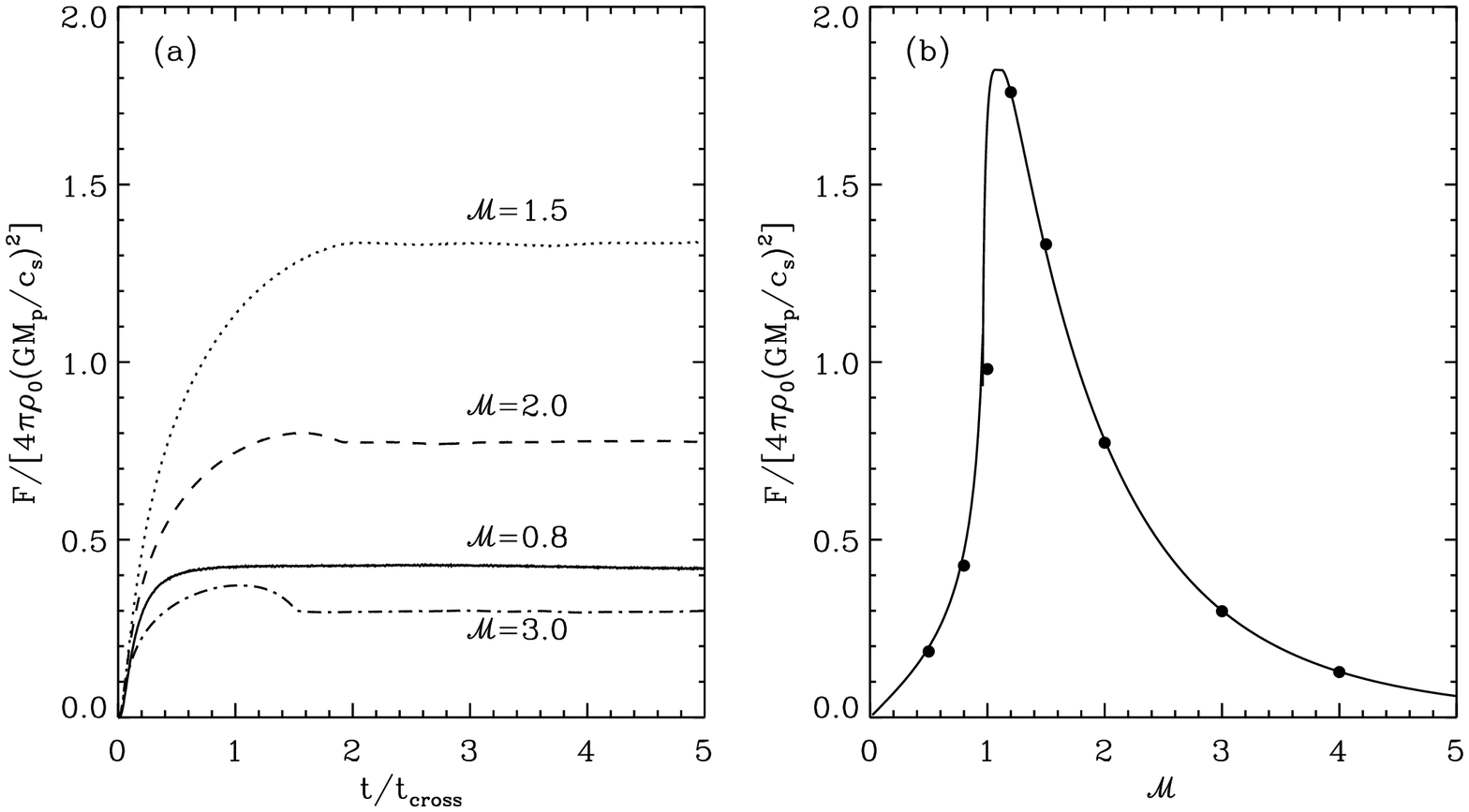}
  \caption{\label{fig:lin_df}
  (\emph{a}) Temporal evolution of the dimensionless DF force for
  $\R=0.02$, $\B=5\times 10^{-4}$ ($\A=0.01)$, and
  $\M=0.8$, 1.5, 2.0, and 3.0. The drag force reaches
  a steady-state value within $\sim 2 \tcr$. (\emph{b}) The mean
  drag force (solid circles), averaged over $t/\tcr= 2$--5, against
  the Mach number. The standard deviation is smaller than the size of the
  circles.
  The solid line plots the linear force formula (eq.\ [\ref{eq:f_pt_linear}])
  with $V_pt=2R_p$ and $\rmin/\rs=1.5+0.7(\M-1.8)^2$ in good agreement
  with the numerical results.
  }
\end{figure*}

We integrate the basic equations of ideal hydrodynamics using a
modified version of the ZEUS code \citep{sto92}, parallelized on a
distributed-memory platform.  Our simulations domain is a
rectangular box which spans the region $-10\Rp \leq x,y,z \leq
10\Rp$. We construct a logarithmically spaced Cartesian grid with
$512^3$ zones, with spacings of $0.0075\Rp$, $0.018\Rp$, and
$0.11\Rp$ at $\vert x\vert/\Rp=0$, 1, and 10, respectively. For some
selected parameters, we have also run low-resolution simulations
with $256^3$ zones and checked that the resulting drag forces agree,
within a few percents, with those from the high-resolution models.
We adopt the outflow boundary conditions at all the boundaries and
run the simulations typically up to $15\tcr$. The simulation
outcomes are insensitive to the box size and the boundary conditions
we adopt since it takes about $\sim 20\tcr$ for sound waves to
travel across the simulation box, while the drag forces saturate
typically within $\sim 0.5\torb \sim(2-3)\tcr$.

\section{Linear Cases} \label{sec:linear}

The time-dependent linear-perturbation analyses for the wakes of
low-mass perturbers usually consider a point-mass object with
vanishing $\rs$ in equation (\ref{eq:Plummer}), which in turn
requires to introduce the cut-off radius $\rmin$ in the linear DF
force formula (e.g, \citealt{jus90,ost99}; KK07).  On the other
hand, one has to adopt a non-zero value for the softening radius in
numerical simulations, making the DF force dependent on $\rs$ in the
linear regime. To explore how the perturber mass affects the drag
force in comparison with the linear theory, therefore, it is
necessary to find a proper relationship between $\rs$ and $\rmin$
that makes the numerical results consistent with the analytic
predictions when $\B\ll1$.  For this purpose, we present in this
section the results of numerical simulations for low-mass perturbers
with $\B=5\times 10^{-4}$ ($\A=0.01$) and $\R=0.05$ as functions of
$\M$.

Figure \ref{fig:lin_wake} shows the snapshots of density wakes on
the $x$--$y$ plane at $t/\torb=1$ for $\M=0.8$, 1.5, 2, and 3 cases.
The perturber in each frame is moving along the white circle in the
counterclockwise direction. For subsonic perturbers, the circular
orbit makes the density wakes curved along their orbits that would
otherwise remain symmetric with respect to the line of motion. For
supersonic models, the perturber is able to overtake its own wake,
creating a strong trailing tail bounded by Mach waves (KK07). As the
Mach number increases from unity, the opening angle of the head of
the curved Mach cone deceases, while the tail thickens. The dotted
lines for the supersonic models plot equation (B2) of KK07 for the
shape of the tail\footnote{Equation (B2) in KK07 reduces to
$R/R_p=\M^{-1}s$ for $R/R_p \gg1$, with $s$ being the azimuthal
angle, suggesting that the wake tail at large radii resembles an
Archimedean spiral.}, which are in excellent agreement with the
outer boundaries of the tails produced in the simulations.
Nevertheless, the finite perturber size in the latter makes the tail
boundaries broader compared to the point-mass counterparts with
$\rs=0$.

For a given density wake $\rho(\vct{x},t)$, it is straightforward to
calculate the gravitational drag force on the perturber by evaluating
the integral
\begin{equation}\label{eq:df}
  F(t)=\int
  \frac{G \MP (|\vct{x}-\vct{x}_p|) \rho(\vct{x},t)
        (\vct{x}-\vct{x}_p)
\cdot \hat \varphi
        }%
       {|\vct{x}-\vct{x}_p|^3}
 \ d^3\vct{x},
\end{equation}
where $\MP(r)= M_p r^3/(r^2 + \rs^2)^{3/2}$ is the mass distribution
of the Plummer sphere and $\hat \varphi={\vct{\hat z}} \times
{\vct{\hat x_p}}$ is the unit vector in the azimuthal direction.
Figure \ref{fig:lin_df}\emph{a} shows the temporal evolution of
$F(t)$ for the cases shown in Figure \ref{fig:lin_wake}. The
friction force saturates to a constant value in less than $\sim
2\tcr$; the amplitudes of fluctuations in $F$ are less than $1\%$ of
the mean values. This is unlike in the linear-trajectory models
where the DF force on a supersonic perturber increases
logarithmically with time, since the whole density wake growing in
size with time is located behind the perturber. The region of
influence in the circular-orbit models expands at a sonic speed from
the orbit center. Consequently, the far-field wake in these models
is more or less spherically symmetric and thus does not contribute
to the DF force much. Figure \ref{fig:lin_df}\emph{b} plots as solid
circles the steady-state drag forces against $\M$ for low-mass
perturbers with $\B=5\times 10^{-4}$ ($\A=0.01$) and $\R=0.05$. The
solid line represents the fit using equation (\ref{eq:f_pt_linear})
with
\begin{equation}\label{eq:rmin}
\rmin/\rs=1.5+0.7(\M-1.8)^2,
\end{equation}
which is in good agreement with the numerical results.
In what follows, we will use equation (\ref{eq:rmin}) as the cut-off
radius when we compare the numerical DF forces with the analytic estimates.

\begin{figure*}
  \epsscale{0.9}
  \plotone{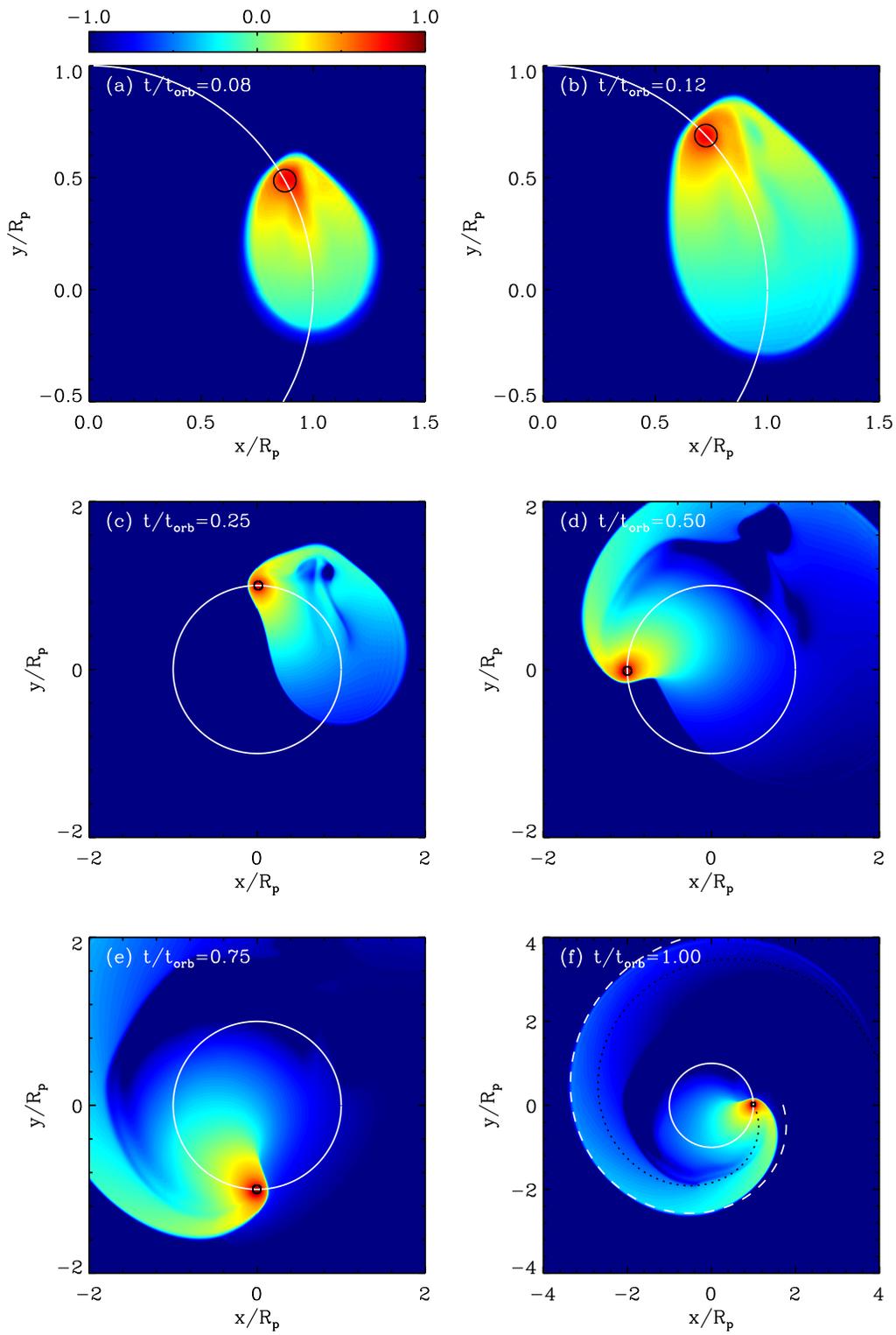}
  \caption{\label{fig:den_sup}
  Density shapshots on the $x$--$y$ plane for a supersonic model with
  $\M=2$, $\R=0.05$, and $\B=0.5$ ($\A=10$).
  The size of the region displayed differs from panel to panel for clarity.
  The white circle or part of it in each panel denotes the orbit, while
  the small black circle marks the softening radius of the perturber.
  In (\emph{f}), the black dotted line plots equation (B2) of KK07 for the linear
  tail, while the white dashed line that matches the outer edge of
  the nonlinear tail is drawn by rotating the former by $75^\circ$ in the counterclockwise
  direction.
  Colorbar labels $\log (\rho/\rho_0-1)$.
  See text for detail.
   }
\end{figure*}

\section{Nonlinear Cases} \label{sec:nonl}

\subsection{Wake Evolution}\label{sec:evol}

\begin{figure*}
  \epsscale{0.8}
  \plotone{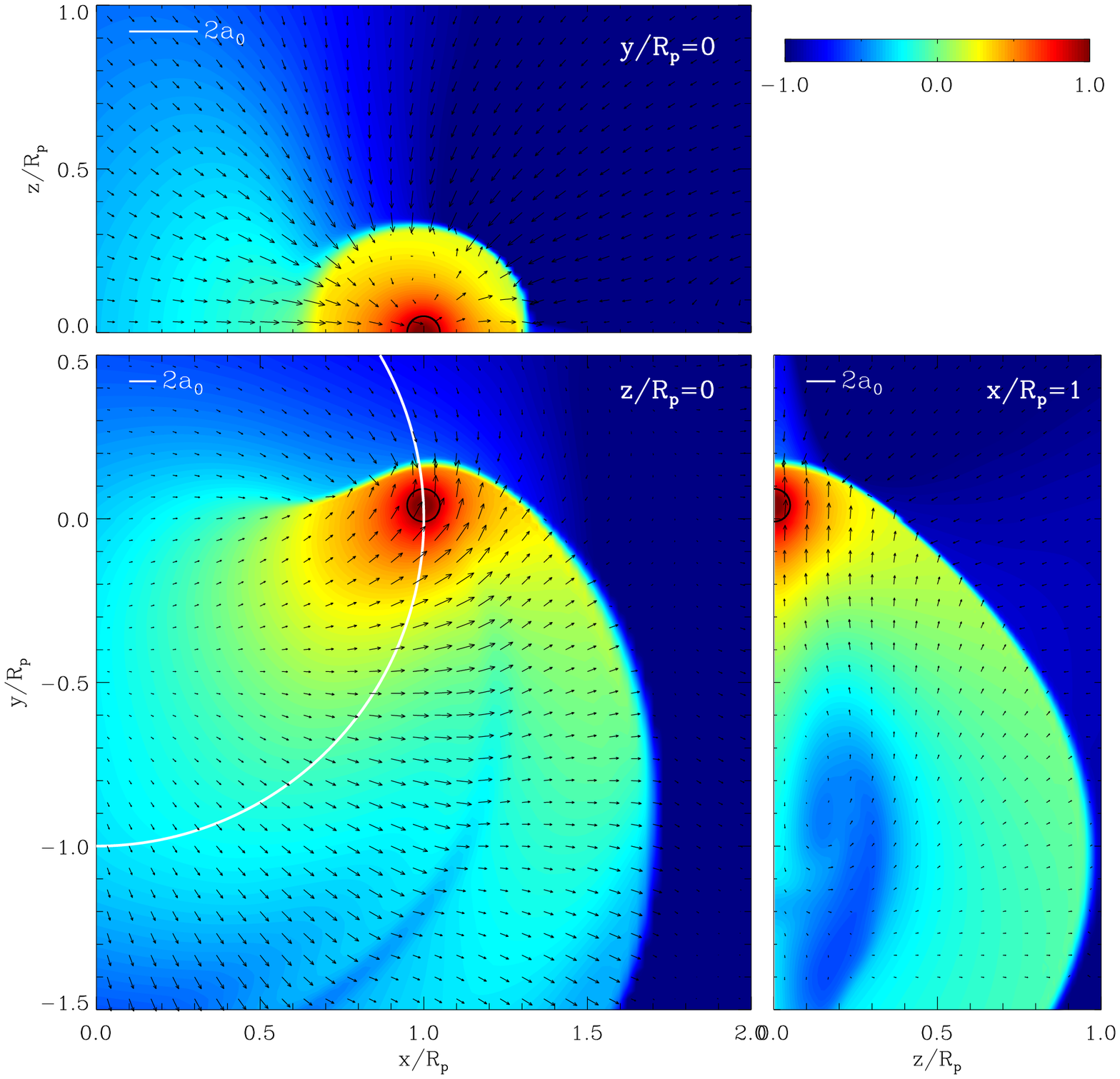}
  \caption{\label{fig:den_vel}
  Quasi-steady distributions of the density wake and velocity field,
  seen in the inertial frame, on the
  $x/R_p=1$ (\emph{bottom right}), $y=0$ (\emph{top left}), and
  $z=0$ (\emph{bottom left}) planes
  for a supersonic model with
  $\M=2$, $\R=0.05$, and $\B=0.5$ ($\A=10$) at $t/\torb=2$.
  The small black circle marks the softening radius of the perturber,
  while the white curve in the bottom left panel draws the perturber orbit.
  The length of the line segment shown in the upper left corner
  of each panel, corresponding to twice the sound speed, measures
  the size of the velocity vectors.
  Colorbar labels $\log (\rho/\rho_0-1)$.}
\end{figure*}

We now focus on the cases with very massive perturbers. Figure
\ref{fig:den_sup} plots the snapshots of logarithmic density on the
$x$--$y$ plane for our standard model with $\R=0.05$, $\M=2$, and
$\B=0.5$ ($\A=10$).  In each panel, the white line draws the
perturber orbit, while the small circle in black marks the perturber
location with its size corresponding to the softening radius $\rs$.
The perturber is initially located at $(x,y)=(R_p,0)$ and orbits in
the counterclockwise direction. Note that the size of the region
displayed in Figure \ref{fig:den_sup} differs from panel to panel
for clear presentation. The evolution of the density wake at early
phase is qualitatively similar to the linear-trajectory cases
described in KK09. An introduction of the perturber at $t=0$
provides the background gas with strong perturbations that readily
develop into a bow shock.  The incident gas along the line of motion
is shocked and gathered near the perturber.  This sets up a strong
pressure gradient in the region between the shock and the perturber,
which in turn causes the shock to slowly move away from the
perturber in the upstream direction until it reaches an equilibrium
position. Some gas flowing with a non-zero impact parameter is
gradually pulled by the perturber, adding to the density at the rear
side (Fig.\ \ref{fig:den_sup}$a$). Due to the strong gravity, the
material piled up at the backside is pulled back toward the
perturber, creating a strong counterstream along the line of motion
as well as a surrounding vortex ring (see KK09).

For the linear-trajectory cases, KK09 has shown that the
counterstream slows down due in part to the strong pressure gradient
established near the perturber and in part to a buoyant expansion of
the vortex ring in the lateral direction.  The interaction of the
counterstream with the shocked gas causes the bow shock to move back
and forth around an equilibrium position. During this time, the
vortex ring also undergoes an overstable oscillation whose amplitude
grows secularly with time.  When the vortex ring moves out beyond a
half of the Hoyle-Lyttleton radius $r_{\rm HL} = 2GM_p/V_p^2$, it
becomes less gravitationally bound and soon swept downstream by the
ram pressure of the incident gas, leaving a near-hydrostatic
envelope surrounding the perturber. This is how a density wake
enters into a quasi-steady state in the linear-trajectory cases. For
the circular-orbit cases, however, the line of perturber motion
keeps changing with time. With the perturber displaced from the
direction of the counterstream, the counterstream is virtually
unhindered and travels almost straight in the direction tangent to
the orbit, producing the protruding region in the wake at $x \approx
y\approx 0.85R_p$ shown in Figure \ref{fig:den_sup}$b$ .
Consequently, the vortex ring manifesting its presence by low
density at $x/R_p=0.6$--$0.9$ and $y/R_p=1.1$--$1.2$ in Figure
\ref{fig:den_sup}$c$ that was carried by the counterstream decouples
from the perturber, promptly leaving a quasi-static density
distribution around the perturber. Note that the time to form a
near-hydrostatic envelope is about $\sim 0.5\torb \approx 30
\rs/\cs$, which is shorter than in the linear-trajectory cases by
about an order of magnitude.

\begin{figure*}
  \epsscale{1}
  \plotone{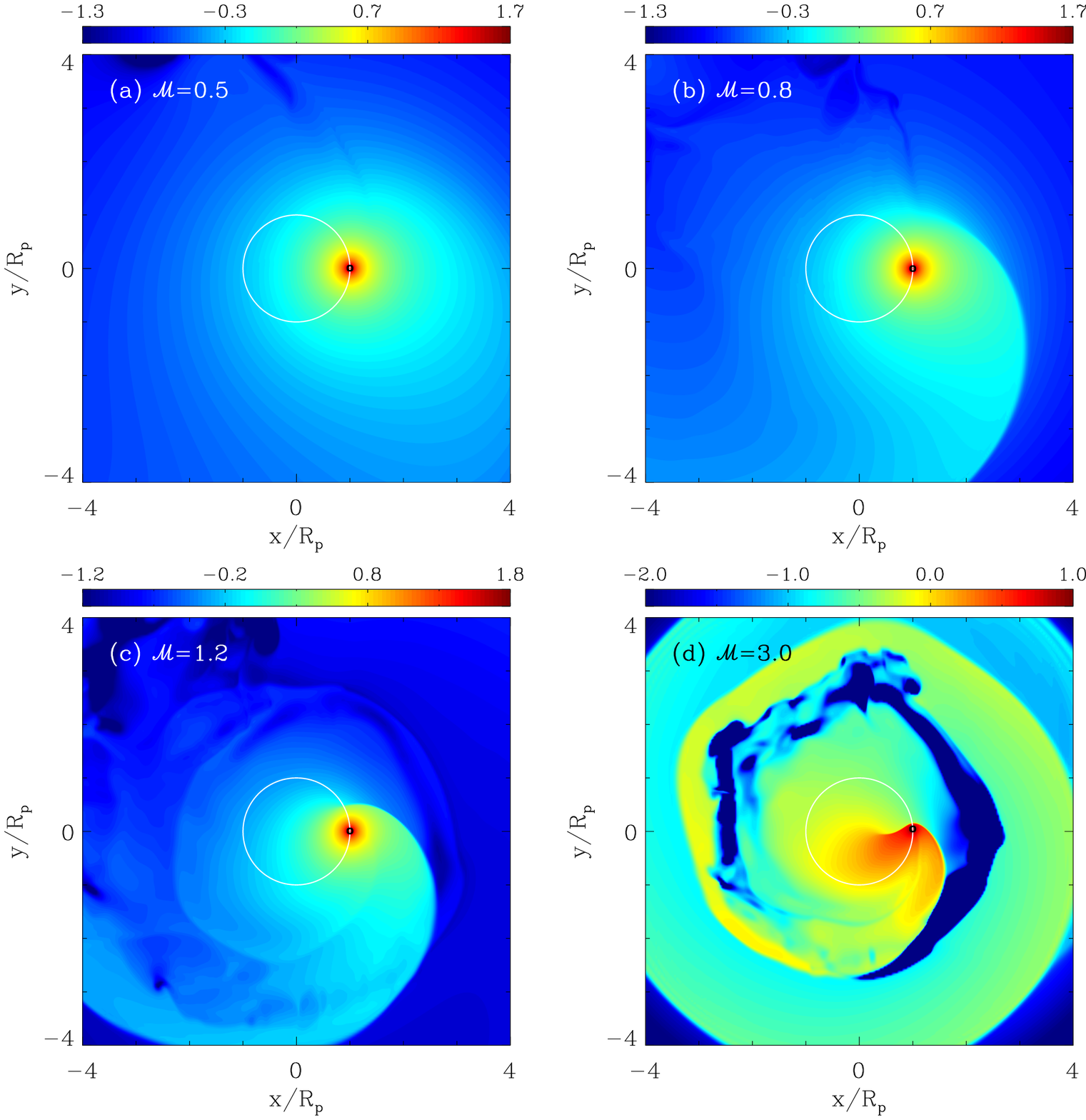}
  \caption{\label{fig:non_wake}
   Distributions of quasi-steady density wakes with differing $\M$
   on the $x$--$y$ plane, with the perturber located at
   $(x,y,z)=(R_p,0,0)$.  All the models have $\R=0.05$ and $\B=1$.
   The white circle in each panel shows the orbit.
   Colorbar labels $\log (\rho/\rho_0-1)$.
   }
\end{figure*}

While the density distribution near the perturber is nearly
spherical, a part of the wake outside the orbit is pushed radially
outward by the counterstream, developing a trailing tail on the
$x$--$y$ plane. As the counterstream weakens, the low-density
regions associated with the vortex ring first merge together and are
then stretched along the orbit as the perturber continues the
orbital motion (Fig.\ \ref{fig:den_sup}$d,e$).  In the low-mass
perturber cases with $\A \ll1$, the wake tail is created by the
overlapping of the Mach cone and the sonic sphere (KK07). In the
nonlinear cases, however, it is the low-density regions produced by
the counterstream that separate the tail from the rest of the wake.
The black dotted curve in Figure \ref{fig:den_sup}\emph{f} plots the
shape of the linear tail, as in Figure \ref{fig:lin_wake}, which
does not match the outer edge of the nonlinear tail in our fiducial
model. Instead, the latter is well matched by rotating the former by
$75^\circ$ in the counterclockwise direction, shown as the dashed
curve.

\subsection{Quasi-steady Density Wakes}\label{sec:wake}

\begin{figure*}
  \epsscale{1}
  \plotone{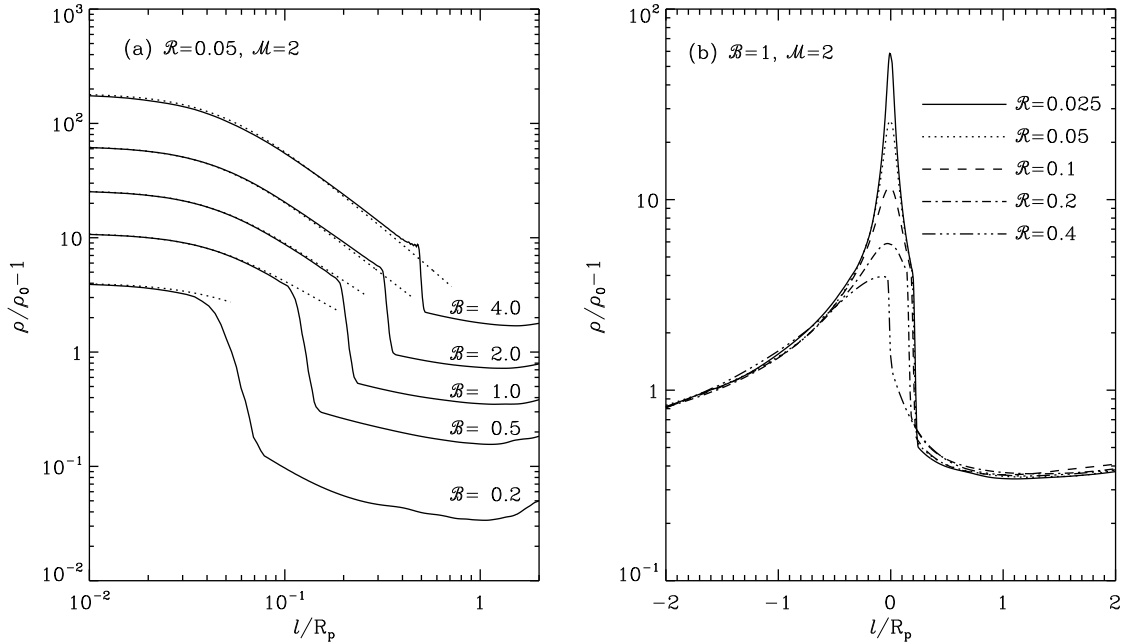}
  \caption{\label{fig:cut_den}
   Density profiles of steady-state wakes as functions of the distance
   $l$ from the perturber center along the azimuthal direction
   for (\emph{a}) models with fixed $\R=0.05$ and varying $\B$
   and (\emph{b}) models with fixed $\B=1$ and varying $\R$. All the
   models have $\M=2$.
   In (\emph{a}), the dotted lines give the respective density
   distributions under the assumption of hydrostatic equilibrium,
   while the numerical results are drawn as the solid lines.
   Note that varying $\R \simlt 0.1$ changes the density
   only in the very central parts of the hydrostatic envelopes.
   }
\end{figure*}

Figure \ref{fig:den_vel} shows the quasi-steady distributions of the
density as well as the velocity field, seen in the inertial frame,
on the $x/R_p=1$, $y=0$, and $z=0$ planes for our fiducial model at
$t/\torb=2$. The colorbar labels the perturbed density in
logarithmic scale. The size of the short line segment shown in the
upper left corner of each panel, corresponding to twice the sound
speed, measures the amplitudes of the velocity vectors. The density
wake consists of a spherical envelope surrounding the perturber, a
detached bow shock, and an extended low-density region at the rear
side.  The envelope is almost hydrostatic, as evidenced by
low-amplitude velocity vectors in the $y=0$ plane to which the
perturber motion is perpendicular instantaneously. The bow shock
standing outside the orbit, in the $z=0$ plane, is just a curved
version of that in the linear trajectory counterpart and extends to
large radii $R=(x^2+y^2)^{1/2}$ in the orbital plane. On the other
hand, the (almost planner) shock located inside the orbit gradually
weakens at small $R$ and terminates at $R/R_p\sim 0.5$--$0.6$ since
the speed of the density wake corotating with the perturber becomes
subsonic there. The low-density regions delineating the tail seen at
$t/\torb=1$ (Fig.\ \ref{fig:den_sup}\emph{f}) are less apparent in
Figure \ref{fig:den_vel} as they diffuse out over time by pressure
gradients. While the velocity just ahead of the perturber is
parallel to the direction of the instantaneous perturber motion, the
direction of gas motions at the perturber location and the immediate
behind it are inclined about $30^\circ$--$50^\circ$ from the
tangential direction. This implies that the gas in the downstream
side retains the information on the direction of perturber motion in
the past.

Figure \ref{fig:non_wake} displays changes in the quasi-steady
density wakes with varying $\M$ on the orbital plane for models with
$\R=0.02$ and $\B=1$ ($\A=20$). In each panel, the perturber located
at $x/R_p=1$ and $y=0$ is moving along the white circle in the
counterclockwise direction. When $\M=0.5$, the density distribution
is circular near the perturber and becomes eccentric away from it,
with the minor axis parallel to the line of motion. Still, there
exists a slight excess of the perturbed density in the rear side,
giving rise to a non-vanishing (but small) DF force. Compared to the
$\M=0.5$ model, the model with $\M=0.8$ has a larger density excess
at the backside. While the perturber is subsonic in this model, a
strong gravitational pull is able to accelerate the background gas
to a near-transonic speed, creating a weak shock outside the orbit.
The shape of the density wake in the $\M=1.2$ model is similar to
that for the $\M=2.0$ model, although the former has a larger peak
density  and a larger distance of the detached bow shock. As $\M$
increases further, both the detached shock distance and the size of
the hydrostatic envelope decrease.  The ring-like low-density gap at
$R/R_p\sim 2$ in Figure \ref{fig:non_wake}\emph{d} is a remnant of
the counterstream that was stretched along the orbit. With large
values of $\M$ and $\B$, the counterstream in this model was so
strong that the deep gap has not been filled in yet.

In order to check whether the density wake near the perturber is in
hydrostatic equilibrium, we plot in Figure \ref{fig:cut_den} the
density profiles of steady-state wakes as functions of distance $l$
measured from the perturber location along the azimuthal direction
in models with $\M=2$. The solid lines in Figure
\ref{fig:cut_den}\emph{a} give the  simulation results for
perturbers with fixed $\R=0.05$ and differing $\B$. The sharp
discontinuity in each curve is due to the presence of the bow shock.
The dotted lines draw the density distributions of the hydrostatic
spheres
\begin{equation}\label{eq:HSE}
\rho(l)=\rho_c
  \left\{1+\frac{(\gamma-1)\cs^2\B}{a_c^2\R}%
       \left[\frac{\rs}{(l^2+\rs^2)^{1/2}}-1\right]%
  \right\}^{1/(\gamma-1)},
\end{equation}
where $\rho_c$ and $a_c$ denote the density and adiabatic sound
speed at $l=0$ (i.e., the perturber center), respectively (KK09).
These are in good agreement with the density profiles of the
steady-state  wakes in our simulations, suggesting that the
envelopes are indeed almost hydrostatic. Figure
\ref{fig:cut_den}\emph{b} shows how the density profile varies with
$\R$.  When $\B=1$, the nonlinear effect is not significant in
models with $\R>0.33$ since they have $\etaA < 1$; the density jump
occurring at $l\sim0$ in the $\R=0.4$ model is more like a Mach wave
rather than a shock. As $\R$ decreases (or $\etaA$ increases), the
density wake becomes increasingly nonlinear, producing a hydrostatic
envelope and a detached bow shock. Note that the shock location is
unchanged as long as $\R \simlt 0.2$ (or $\etaA \simgt 2$). When the
density is highly nonlinear, the strong gravity associated with
small $\rs$ increases the density only in the small regions within
$\simlt (1-2)\rs$.  Since this core region is spherically symmetric,
the resultant DF force becomes independent of $\R$ provided $\etaA
\simgt 2$. We will show this directly in \S\ref{sec:df} below.

\begin{figure}
  \epsscale{1}
  \plotone{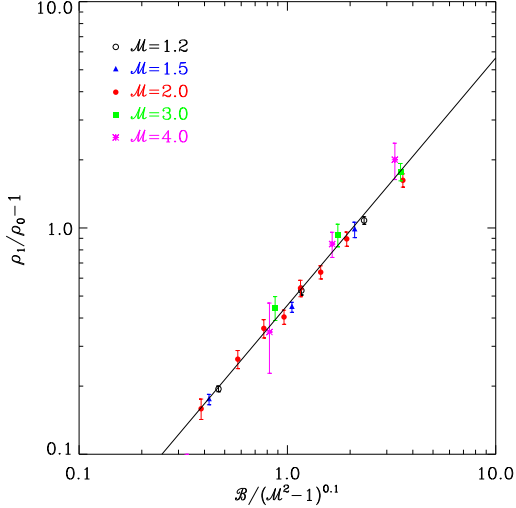}
  \caption{\label{fig:rho1}
  Modified background density $\rho_1$ ahead of the bow shock
  in models with $\R=0.05$.
  Various symbols and errorbars give the means and standard deviations
  of the preshock density in the regions with
  $\Delta l/R_p = 0.5$--$1.5$ from the detached shock.
  The solid line plots the best fit given by equation (\ref{eq:rho1}).
   }
\end{figure}

Unlike in the linear-trajectory cases where a perturber travels
through an undisturbed medium, it is inevitable that a
circular-orbit perturber enters the backside of its own wake before
completing one orbit. In the circular-orbit cases, therefore, the
density ahead of the bow shock differs from the undisturbed density
$\rho_0$ and depends considerably on $\B$, as Figure
\ref{fig:cut_den}\emph{a} illustrates. The enhanced preshock density
can alternatively be interpreted as arising from the response of the
background gas to the orbit-averaged gravitational potential of the
perturber $\langle \PhiP\rangle \equiv \int_0^{\torb} \PhiP(t)
dt/\torb$, which is centered at the orbit center. In this sense, the
enhanced preshock density $\rho_1$ acts as a modified background
density to a circular-orbit perturber with  the gravitational
potential $ \PhiP - \langle \PhiP\rangle$. To quantify $\rho_1$, we
take a spatial average of the preshock density in quasi-steady state
over the region with $(l-\delta)/\Rp=0.5$--$1.5$ along the orbit,
where $\delta$ denotes the location of the detached shock. Figure
\ref{fig:rho1} plots the resulting $\rho_1$ for models with
$\R=0.05$, with errorbars representing the standard deviations. The
solid line is the best fit
\begin{equation}\label{eq:rho1}
\frac{\rho_1}{\rho_0}= 1 + \frac{0.46 \B^{1.1}}{(\M^2-1)^{0.11}},
\end{equation}
showing that the change in the background density to circular-orbit
perturbers is almost linearly proportional to $\B$.
Since $\rho_1$ is quite insensitive to $\M$, the $\M$-dependence
of $\rho_1$  in equation (\ref{eq:rho1}) can be ignored.

\subsection{Detached Shock Distance}\label{sec:delta}

\begin{figure*}
  \epsscale{1}
  \plotone{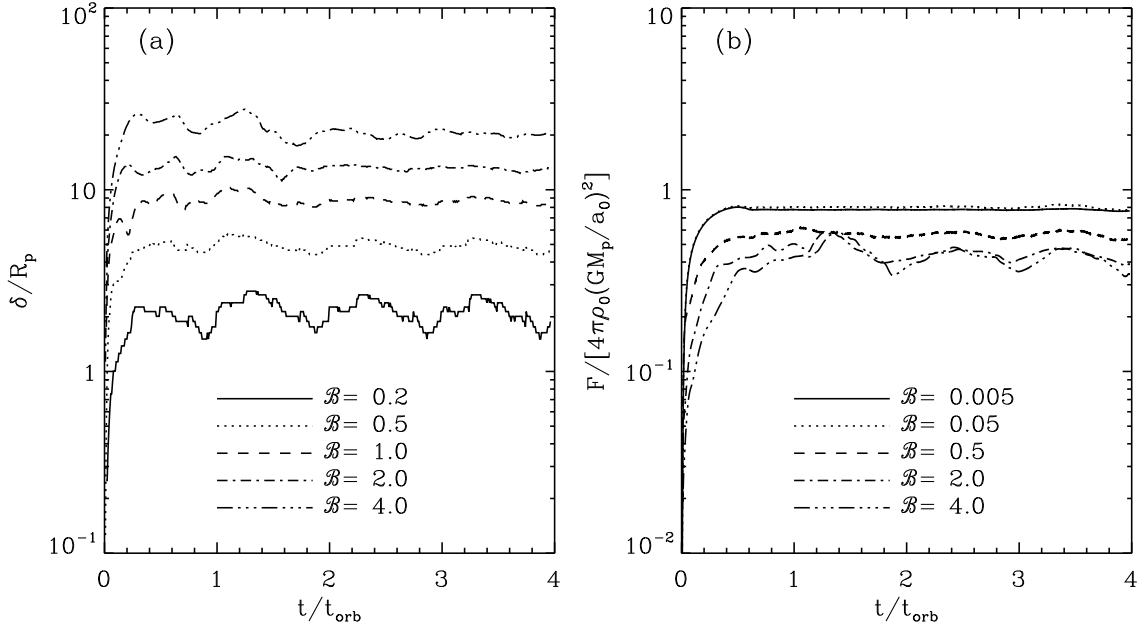}
  \caption{\label{fig:delta_time}
  Temporal evolution of
  (\emph{a}) the detached shock distances $\delta$ and
  (\emph{b}) the DF forces $F$ for models $\R=0.05$ and $\M=2$.
  Both $\delta$ and $F$ increase rapidly with time initially and
  then saturate at $t/\torb\sim0.5$ to equilibrium values with
   some fluctuations. The dimensionless drag force decreases with
   $\B$ for $0.05<\B<2$ due to the nonlinear effect.
   }
\end{figure*}

In laboratory experiments and hydrodynamic theories, supersonic
flows over a blunt-body generate a detached bow shock when the
maximum angle allowed in the postshock flows is smaller than the
nose angle (e.g., \citealt{lie57,shu92}).  Although a perturber in
our models does not hold any defined surface and merely provides
gravitational perturbations to the background gas, a spherical
envelope formed around it nevertheless sends off pressure waves into
the upstream direction, acting as an obstacle to the incoming gas.
Since the nose-angle of a spherical body is $90^\circ$, the shock
should be detached in our nonlinear supersonic models. Figure
\ref{fig:delta_time}\emph{a} plots the time evolution of the
detached shock distance $\delta$ for some selected models with
$\R=0.02$ and $\M=2$. Note that $\delta$ initially increases rapidly
and then saturates at $t/\torb\sim 0.5$ to an equilibrium value with
some fluctuations. The quasi-steady value of $\delta$ becomes larger
as the perturber mass increases.

\begin{figure}
  \epsscale{1}
  \plotone{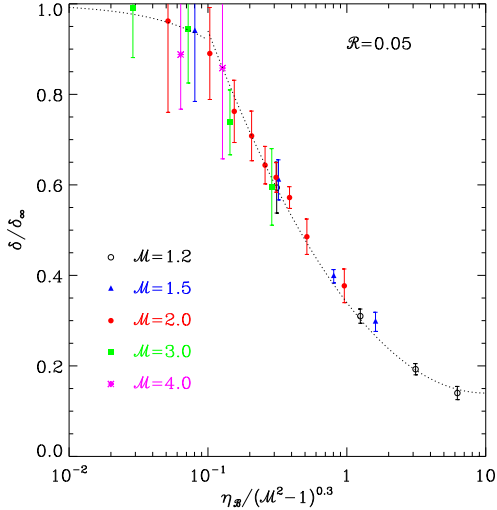}
  \caption{\label{fig:delta}
  Quasi-steady values of the detached shock distances $\delta$ of
  circular-orbit perturbers relative to those, $\delta_\infty$,
  of linear-trajectory perturbers.
  Various symbols and errorbars give the means and standard deviations
  of $\delta/\delta_\infty$ over $t/\torb=1$--$3$.
  The dotted curves,  expressed as equation (\ref{eq:del_cir}),
  are the best fits to the numerical results.
   }
\end{figure}

For massive perturbers moving on a straight-line trajectory,
KK09 found that the detached shock distance is given empirically by
\begin{equation}\label{eq:del_lin}
\delta_\infty = \rs\etaA = \frac{GM_p}{\cs^2(\M^2-1)}.
\end{equation}
This can be interpreted by the balance between the postshock thermal
energy ($\propto \cs^2 \M^2$) and the gravitational potential energy
($\propto -GM_p/\delta$) at the shock location for strong shocks
($\M\gg1$). To study how the circular obit affects the standoff
distance of the shock, we take a time average of $\delta(t)$ over
$t/\torb\sim 1$--$3$ and plot $\delta/\delta_\infty$ in Figure
\ref{fig:delta} as various symbols. Again, errorbars indicate the
standard deviations of $\delta(t)$ during this time interval.   The
numerical results show that the detached shock distances in the
circular orbits are in general smaller than those in the
linear-trajectory cases.  This is likely because the continuous
change in the direction of perturber motion effectively reduces the
forward momentum of the incident gas near the perturber, as
illustrated  by the spatially-varying velocity field in the wake
shown in Figure \ref{fig:den_vel}. In this case, the ram pressure of
the incoming flow is able to push the shock front toward the
perturber, making $\delta$ smaller. The modified background density
and sound speed discussed in \S\ref{sec:wake} might affect $\delta$
as well. We fit our results using
\begin{equation}\label{eq:del_cir}
\frac{\delta}{\delta_\infty} =
\left\{\begin{array}{ll}
  1 - 0.8 \Lambda , & \;\;\;\;\;\;\;\;\;\,\Lambda < 0.1,  \\
  0.14 + 0.2 ( 1- \log\Lambda )^{2}, & 0.1 < \Lambda <10 ,
\end{array}\right.
\end{equation}
with $\Lambda \equiv \etaB/(\M^2-1)^{0.3}$, which is plotted as
dotted lines. As expected, $\delta/\delta_\infty \rightarrow 1$ as
$\Lambda\rightarrow 0$ corresponding to $\Rp\rightarrow\infty$,
although models with $\Lambda \simlt 0.1$ are weakly nonlinear and
have $\delta$ comparable to or less than $\rs$. Since
$\delta_\infty/R_p=\etaB$,  equation (\ref{eq:del_cir}) indicates
that $\delta$ increases with $\Lambda$, although its increasing rate
slows down as $\Lambda \rightarrow 10$ where $\delta$ becomes
comparable to $R_p$.

\subsection{Drag Force}\label{sec:df}

For our nonlinear models, we calculate the DF force on a perturber
due to its own induced wake.  Figure \ref{fig:delta_time}\emph{b}
illustrates the temporal changes of the dimensionless drag force
$\I\equiv F / [4\pi \rho_0 (GM_p/a_0)^2]$ in models with $\M=2$ and
$\R=0.02$ and varying $\B$. The drag force reaches a quasi-steady
value in less than one orbit, although it exhibits some fluctuations
for large $\B$. This is markedly different from the cases of
straight-line trajectories where the drag force increases
logarithmically with time. The quasi-steady value of $\I$ is almost
independent of $\B$ when $\B \simlt 0.05$, decreases with $\B$ for $
0.05\simlt \B \simlt 2$, and becomes again insensitive to $\B$ when
$\B \simgt 2$. The behavior of $\I$ with $\B$ is a consequence of
the circular orbit combined with the nonlinear effect. The size of a
hydrostatic envelope in a nonlinear density wake, as measured by the
detached shock distance, becomes larger with increasing $\B$, which
makes the drag force smaller owing to the front-back symmetry in the
wake. At the same time, the circular orbit draws the background
material toward the orbit  center and increases the density in the
preshock region, which in turn tends to increase the drag force. For
models with $\B \simlt 2$ (so that $\rho_1/\rho_0 \simlt 2$), the
nonlinear effect dominates and $\I$ decreases with $\B$. When
$\B\simgt2$, on the other hand, the enhancement of the background
density becomes significant enough to compensate for the nonlinear
effect, making $\I$ unchanged with $\B$. As we will show below, the
DF force scales well with $\B$ when $\rho_1/\rho_0$ is taken into
account.

\begin{figure}
  \epsscale{1}
  \plotone{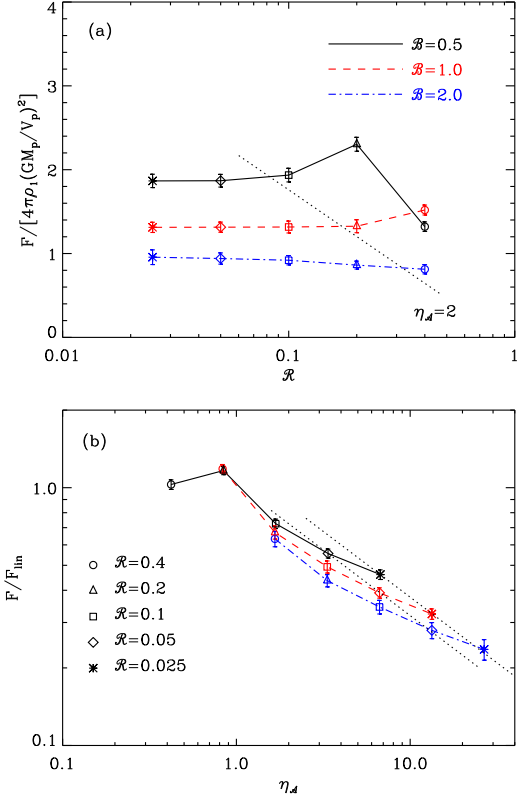}
  \caption{\label{fig:df_rs}
   Normalized DF forces averaged over $t/\torb=2$--$4$
   for models with $\M=2$ and various $\B$ and $\R$
   as functions of (\emph{a}) $\R$ and (\emph{b}) $\etaA$.
   The dotted line corresponding to $\etaA=2$ in (\emph{a})
   separates the domain into two parts such that the models at
   the lower-left part are fully nonlinear.
   Dotted lines in (\emph{b}) have a slope of $-0.5$ and
   connect the numerical results for $\R=0.05$ and 0.025
   models quite well.
   }
\end{figure}

Compared to the linear-trajectory cases, the circular orbit
introduces an additional length scale, $R_p$, so that it is
interesting to explore the dependence of the drag force upon $\R$.
Figure \ref{fig:df_rs}\emph{a} plots as various symbols and
errorbars the mean values and standard deviations of the drag forces
averaged over $t/\torb=2$--4 for models with $\M=2$.  Note that the
drag force is normalized by $\rho_1$ instead of $\rho_0$. The dotted
line marked with $\etaA=2$ demarcates the parameter space into two
parts such that the models shown in the lower-left region are fully
nonlinear with $\etaA>2$, while those in the upper-right region are
in the linear or weakly nonlinear regime. The normalized drag force
on weakly-nonlinear perturbers with $0.7<\etaA<2$ is usually larger
than that on low-mass perturbers with $\etaA<0.7$ because the
density wakes in the former are distributed closer to the perturber
(KK09). In highly-nonlinear models with $\etaA>2$ (or $\etaB>2\R$),
on the other hand, the DF force becomes essentially independent of
$\R$. This is of course because the change in $\rs$ affects the wake
only within a small region close to the perturber where the density
is spherically symmetric, making a negligible contribution to the
drag force (see \S\ref{sec:wake}).

\begin{figure}
  \epsscale{1}
  \plotone{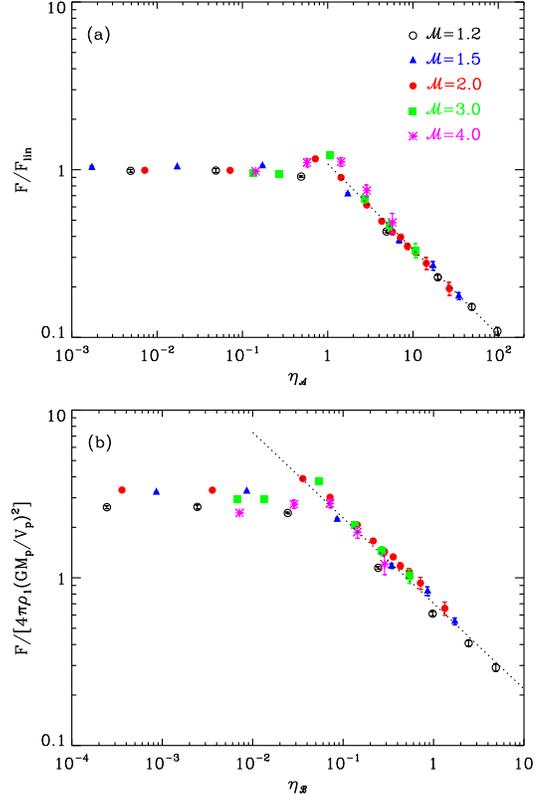}
  \caption{\label{fig:df_Non}
   (\emph{a}) DF forces on supersonic perturbers with $\R=0.05$
   normalized by the linear drag force as a function of $\etaA$.
   Various symbols and their errorbars give the means and standard
   deviations of $F/\Flin$ for $t/\torb=2$--$4$.
   The dotted line indicates a slope of $-0.5$.
   (\emph{b}) The same data as in (\emph{a}) but with different
   normalization and as a function of $\etaB$.
   The dotted line shows the best fit to the numerical results for
   $\etaB>0.1$ and is given by equation (\ref{eq:nonlinfit}).
   }
\end{figure}

Figure \ref{fig:df_rs}\emph{b} plots $F/\Flin$ versus $\etaA$
for the same models shown in Figure \ref{fig:df_rs}\emph{a}.
Note again that $\rho_0$ in equation (\ref{eq:f_pt_linear})
for $\Flin$ is replaced by $\rho_1$.
For given $\B$ (so that increasing $\etaA$ corresponds to
decreasing $\R$), $F/\Flin$ decreases logarithmically with
increasing $\etaA > 2$. This is because $\etaA\propto \R^{-1}$ for
fixed $\B$ and $\M$, leading to $\Flin \propto \ln\etaA$,
while $F$ is constant for $\etaA>2$.
For fixed $\R$ (so that increasing $\etaA$ corresponds to increasing $\B$),
$F/\Flin \propto \etaA^{-0.5}$, as illustrated by two dotted lines
connecting the numerical results for $\R=0.05$ and 0.025, respectively.
Figure \ref{fig:df_Non}\emph{a} plots $F/\Flin$ for all of our
supersonic models with $\R=0.05$.  Obviously,
$F/\Flin\approx 1$ for low-mass perturbers with $\etaA<0.7$.
The dotted line is the best fit
\begin{equation}\label{eq:linfit}
\frac{F}{\Flin} = 1.1 \etaA^{-0.5}, {\rm\;\; for\;}\etaA>2,
\end{equation}
to the numerical results for nonlinear, circular-orbit perturbers,
which is very similar to equation (\ref{eq:KK09}) for massive,
linear-trajectory perturbers.

In order to apply equation (\ref{eq:linfit}) to estimate the decay
timescale, one needs to specify the softening radius which is not
directly observable in most astronomical systems. In addition,
equation (\ref{eq:linfit}) is applicable only when $\R=0.05$ since
$F$ on highly massive perturbers is independent of $\rs$, while
$\Flin$ diverges as $\rs\rightarrow0$. For practical use in various
situations, therefore, it is desirable to have an expression of $F$
that does not rely on $\rs$. For this purpose, we plot in Figure
\ref{fig:df_Non}\emph{b} the dimensionless DF forces as a function
of $\etaB$ for all supersonic models with $\R=0.05$. The dotted line
is our fit to the nonlinear part using
\begin{equation}\label{eq:nonlinfit}
F = \frac{4\pi\rho_1 (GM_p)^2}{V_p^2}
\left(\frac{0.7}{\etaB^{0.5}}\right), {\rm\;\; for\;}\etaB> 0.1,
\end{equation}
with $\rho_1$ given by equation (\ref{eq:rho1}). For the models with
$\etaB<0.1$, equations (\ref{eq:f_pt_linear}) and (\ref{eq:rmin})
with $V_pt =2R_p$ provide a good estimate of $F$ that depends on
$\rs$.

\section{Summary and Discussion}\label{sec:diss}

DF due a gaseous medium may play a central role in removing angular
momentum from astronomical objects in orbital motions, causing them
to spiral in toward the orbit center. In this paper, we use
three-dimensional hydrodynamic simulations to explore the
gravitational wake and the associated drag force on a very massive
perturber with mass $M_p$ moving at speed $V_p$ on a circular orbit
with radius $R_p$. This work extends our previous studies that
considered low-mass, circular-orbit perturbers (KK07) and high-mass,
linear-trajectory perturbers (KK09).  The background medium is
assumed to be uniform with density $\rho_0$ and the sound speed
$\cs$, adiabatic with index $\gamma=5/3$, non-self-gravitating,
non-rotating, and unmagnetized. The perturber is represented by a
Plummer sphere with softening radius $\rs$.  Our models are
characterized completely by three dimensionless parameters:
$\R=\rs/R_p$, $\M=V_p/\cs$, and $\B=GM_p/(\cs^2R_p)$. We run a total
of 65 models that differ in $\R$, $\M$, and $\B$, and obtain the
dependence of the drag force on these parameters.

For our supersonic models, the density wake of a massive perturber
reaches a quasi-steady state typically in $\sim 0.5\torb$, about 10
times faster than the linear-trajectory cases. A quasi-steady
density wake consists of a spherical envelope surrounding the
perturber, a detached bow shock in the upstream direction, and an
extended low-density region in the rear side. The envelope is almost
hydrostatic, providing a negligible contribution to the DF force.
Compared to the linear-trajectory cases, an orbit-averaged
gravitational potential of a circular-orbit perturber is able to
gather the background gas toward the orbit center, effectively
increasing the background density ahead of the bow shock. Equation
(\ref{eq:rho1}) describes the modified preshock density $\rho_1$,
showing that density enhancement in the preshock region is almost
linearly proportional to $\B$ and depends weakly on $\M$.

Strong perturbations sent off from a massive perturber develop into
a bow shock through which the upstream supersonic flow becomes
subsonic. Since the density wake in a quasi-steady state forms a
pattern that is corotating with the perturber at a constant angular
speed, the incident flow outside the orbit is supersonic to a
perturber with $\M>1$. This naturally produces a bow shock outside
the orbit that is simply a curved version of that in the
linear-trajectory counterpart. Formation of a weak bow shock outside
the orbit is possible even for a subsonic perturber with $\M\simgt
0.8$ for which the incident flow is accelerated to near-transonic
speeds. On the other hand, the motion of the steady-state wake is
subsonic to the gas near the orbit center, even when $\M>1$. This
makes the bow shock gradually weaken at small radii, eventually
terminating somewhere inside the orbit. The detached shock distance
measured along the tangential direction is generally smaller than in
the liner-trajectory counterpart since the continuous change in the
direction of the perturber motion effectively reduces the forward
momentum of the wake; equation (\ref{eq:del_cir}) gives the
algebraic fits to the numerical results for the detached shock
distance.

Unlike in the linear-trajectory cases where the drag force increases
logarithmically with time, the DF force in the circular orbits
approaches a quasi-steady value in less than one orbit. The
quasi-steady drag force is essentially independent of $\R$ as long
as $\etaA>2$ (or $\etaB>2\R$) since the change in $\rs$ modifies the
density wake only within $\sim(1-2)\rs$ from the perturber center,
where the wake is almost spherically symmetric and thus has a
negligible contribution to the drag force. For sufficiently massive
perturbers, the presence of a near-hydrostatic envelope in the
density wake also makes the nonlinear drag force smaller than the
linear estimate. Since the choice of $\rs$ is uncertain in many
practical applications, we fit the nonlinear drag force on a
circular-orbit perturber using equation (\ref{eq:nonlinfit}) that
does not involve $\rs$.

That the nonlinear drag force in circular orbits is insensitive to
the softening radius seems inconsistent with the results of KK09 for
linear-trajectory perturbers.  But, it does not. For the latter
case, $\rs$ is in fact a lone length scale relative to which other
length scales such as the Bondi radius and the distance traveled by
the perturber are measured. It is then difficult to separate the
effect of $\rs$ on the drag force from that of $M_p$ because the
models depend on them only through the dimensionless parameter $\A$.
Had we run simulations by using \emph{dimensional} quantities in
KK09, we would probably have gotten the results that $F \propto
M_p^{1.5}$ independent of $\rs$ for models with $\etaA\gg 1$,
similarly to the results of the present paper. Given that $\Flin
\propto M_p^2 \ln(V_pt/\rs)$, the time-averaged values of $F/\Flin$,
when expressed in terms of $\etaA$ in KK09, somehow picked up the
power-law dependence on $M_p$ more clearly than the weak logarithmic
dependence on $\rs$. The circular orbit breaks the degeneracy
between $\rs$ and $M_p$ by introducing another length scale, $R_p$,
allowing to separate the effect on the DF force of $\rs$ from that
of $M_p$. The similarity between equation (\ref{eq:linfit}) obtained
for fixed $\R$ in the current work and the result of KK09 suggests
that the latter should be interpreted as the dependence of $F$ on
$M_p$ for fixed $\rs$.

Now we compare the results of our idealized models with those from
realistic simulations for the orbital decay of SMBHs at galaxy
centers (e.g., \citealt{esc04,esc05,dot06,dot07,may07}; see also
\citealt{col09} and references therein). While these authors showed
that SMBHs inspiral rapidly in $\sim10^6-10^7$ yrs due to the
gaseous DF force, they also found that the DF force in supersonic
models is $\sim 1.5$ times smaller, and depends on the black hole
mass less sensitively, than the analytical predictions of
\citet{ost99} formula (e.g., \citealt{esc04,esc05}). The
discrepancies between the numerical and analytical results are most
likely due to the nonlinear effect. For instance, one of the models
considered by \citet{esc05} has a black hole with mass $M_p=5\times
10^7\Msun$ and softening radius $\rs=4$ pc moving initially with
speed $V_p\sim100\kms$ on a circular orbit with $R_p=200$ pc through
a medium with sound speed $\cs\sim 70\kms$, corresponding to
$\B=0.2$, $\M=1.4$, and $\R=0.02$. For this model, equation
(\ref{eq:nonlinfit}) compared with equation (\ref{eq:f_pt_linear})
would give $\Flin/F \sim 2$, consistent with the results of
\citet{esc05}. Since the orbital decay timescale is given by $\taud
= M_pV_p/F$, equation (\ref{eq:nonlinfit}) predicts $\taud \propto
M_p^{-0.5}$ for sufficiently massive perturber, which is more
consistent with $\taud \propto M_p^{-0.3}$ inferred from the results
of \citet{esc04,esc05} than the linear prediction $\taud \propto
M_p^{-1}$. The difference between the results of the current paper
and \citet{esc04,esc05} is presumably due to the effects of density
stratification, rotation, self-gravity in the background medium,
which are not considered in the present work.

Another important issue regards the proper choice of the softening
length of a point-mass perturber employed in numerical simulations
for the gaseous DF. As mentioned in \S\ref{sec:intro}, the limited
numerical resolution commonly requires to take $\rs$ a few orders of
magnitude larger than the realistic size. Are such large values of
$\rs$ acceptable in calculating the orbital decay time accurately?
Based on our numerical results, the answer is yes, provided
$\etaA>2$ (or $\etaB>2\R$). Otherwise, a large value of $\rs$ would
cause the gravity near the black hole to be reduced significantly.
The resulting density wake would then remain in the linear regime,
making the drag force overestimated considerably. For the decay of
SMBHs,  the models considered in \citet{esc04,esc05} have
$\etaA\sim10$, well above the required lower limit. Some models with
$M_p\sim10^6-10^7\Msun$ in \citet{dot07} and \citet{may07} have
$\etaA\sim1$, so that their choices of $\rs\sim0.1-0.2$ pc are
marginally acceptable. For the evolution of a main-sequence
companion with $M=0.4\Msun$ in a common-envelope binary,
\citet{san98} took $\rs\sim2.3\times 10^{11}$ cm, about 10 times
larger than the real size. Since this corresponds to $\etaA\sim20$,
our results suggest that the softening radius did not affect the
decay time in their numerical simulations.

Finally, we remark on the several assumptions made in the current
work. First, we consider a single perturber moving on a circular
orbit. If the resulting drag force (eq.\ [\ref{eq:nonlinfit}]) is to
be applied to the decay of double black holes, one has to take into
account the drag force from the wake of its companion located at the
other side of the orbit, as well. \citet{kim08} found that for
equal-mass perturbers, the ratio of the DF force from the companion
wake to that from its own induced wake is about $\sim10-50\%$ for
supersonic cases, depending on the Mach number, when the perturbers
have so low masses that the wakes are in the linear regime.  The
nonlinear effect on massive perturbers has yet to be explored.
Second, the current study assumes an adiabatic equation of state
with index $\gamma=5/3$. It is not well known what the most
appropriate value of $\gamma$ is in the background medium, but it
would be $\gamma=5/3$ if the gas is fully ionized or $\gamma=7/5$
for preferentially molecular gas, under the assumption that
radiative heating and cooling are unimportant. The numerical results
of \citet{may07} indicate that the orbital decay is more effective
for a softer equation of state, suggesting that the DF force may
depend sensitively on $\gamma$. Third, while we consider a static
background medium with uniform density, it is more likely to be
stratified in real situations. For a collisionless background,
\citet{jus05} found that a density gradient induces an additional
drag force in the lateral direction of the perturber motion,
amounting to about 10\% of the drag in the backward direction. We
also have ignored the rotation, self-gravity, buoyancy, and
turbulent motions of the background medium in the present paper. It
is interesting to see what effect each of these physical ingredients
makes, which will direct our future research.

\acknowledgments The author is grateful to E.~C.\ Ostriker and H.\
Kim for helpful discussions and the anonymous referees for
stimulating comments. This work was supported by the National
Research Foundation of Korea (NRF) grant funded by the Korean
government (MEST), No. 2009-0063616. Simulations were performed by
using the supercomputing resource of the Korea Institute of Science
and Technology Information through the grant KSC-2009-S02-0008.


\begin{thebibliography}{}
\bibitem[Adams et al.(1989)]{ada89}
   Adams, F.\ C., Ruden, S.\ P., \& Shu, F.\ H.\ 1989, \apj, 347, 959
\bibitem[Barausse(2007)]{bar07} Barausse, E.\ 2007, \mnras, 382, 826
\bibitem[Balbus \& Soker(1990)]{bal90}
  Balbus S.\ A., Soker N., 1990, \apj, 357, 353
\bibitem[Begelman et al.(1980)]{beg80}
  Begelman, M.\ C., Blandford, R.\ D., \& Rees, M.\ J.\ 1980, \nat, 287, 307
\bibitem[Binney \& Tremaine(2008)]{bin08}
 Binney, J., \& Tremaine, S.\ 2008, Galactic Dynamics, 2nd ed.
 (Princeton: Princeton Univ.\ Press)
\bibitem[Chambers(2009)]{cha09}
 Chambers, J.~E.\ 2009, Annu.\ Rev.\ Earth Planet.\ Sci.\ 2009, 37, 321
\bibitem[Chandrasekhar(1943)]{cha43} Chandrasekhar, S.\ 1943, \apj, 97, 255
\bibitem[Colpi \& Dotti(2009)]{col09}
  Colpi, M., \& Dotti, M.\ 2009, to appear on Advanced Science Letters;
  arXiv:0906.4339
\bibitem[Conroy \& Ostriker(2008)]{con08}
  Conroy, C., \& Ostriker, J.\ P.\ 2008, \apj, 681, 151
\bibitem[Cuadra et al.(2009)]{cua09} Cuadra, J., Armitage, P.~J., Alexander, R.~D., \& Begelman, M.~C.\ 2009, \mnras, 393, 1423
\bibitem[Dokuchaev(1964)]{dok64} Dokuchaev, V.~P.\ 1964, Soviet Astron., 8, 23
\bibitem[Dotti et al.(2006)]{dot06} Dotti, M., Colpi, M., \& Haardt, F.\ 2006, \mnras, 367, 103
\bibitem[Dotti et al.(2007)]{dot07} Dotti, M., Colpi, M., Haardt, F., \& Mayer, L.\ 2007, \mnras, 379, 956
\bibitem[Edgar(2004)]{edg04} Edgar, R.\ 2004, New Astro.\ Rev., 48, 843
\bibitem[El-Zant et al.(2004)]{elz04} El-Zant, A.~A., Kim, W.-T., \& Kamionkowski, M.\ 2004, \mnras, 354, 169
\bibitem[Escala et al.(2004)]{esc04} Escala, A., Larson, R.~B., Coppi, P.~S., \& Mardones, D.\ 2004, \apj, 607, 765
\bibitem[Escala et al.(2005)]{esc05} Escala, A., Larson, R.~B., Coppi, P.~S., \& Mardones, D.\ 2005, \apj, 630, 152
\bibitem[Ferrarese \& Ford(2005)]{fer05}
  Ferrarese, L., \& Ford, H.\ C.\ 2005, \ssr, 116, 523
\bibitem[Just \& Kegel(1990)]{jus90} Just, A., \& Kegel, W.~H.\ 1990, \aap, 232, 447
\bibitem[Just \& Pe\~narrubia(2005)]{jus05} Just, A., \& Pe\~narrubia, J.\ 2005, \aap, 431, 861
\bibitem[Kim \& Kim(2007)]{kk07} Kim, H., \& Kim, W.-T.\ 2007, \apj,
665, 432 (KK07)
\bibitem[Kim \& Kim(2009)]{kk09} Kim, H., \& Kim, W.-T.\ 2009, \apj, 703,
1278 (KK09)
\bibitem[Kim et al.(2008)]{kim08} Kim, H., Kim, W.-T., \& S{\'a}nchez-Salcedo, F.~J.\ 2008, \apjl, 679, L33
\bibitem[Kim(2007)]{kim07} Kim, W.-T.\ 2007, \apj, 667, L5
\bibitem[Kim et al.(2005)]{kim05} Kim, W.-T., El-Zant, A.~A., \& Kamionkowski, M.\ 2005, \apj, 632, 157
\bibitem[Liepmann \& Roshko(1957)]{lie57} Liepmann, H., \& Roshko, A.\ 1957,
  Elements of Gasdynamics, Galcit Aeronautical Series (New York: Wiley, 1957)
\bibitem[Lubow \& Ida(2010)]{lub10}
  Lubow, S.~H., \& Ida, S.\ 2010, to appear in Exoplanets, ed.\ S.\ Seager, Univ.\ Arizona
  Press; arXiv:1004.4137
\bibitem[Lufkin et al.(1995)]{luf95}
  Lufkin, E. A., Balbus, S.\ A., \& Hawley, J.\ F.\ 1995, \apj, 446, 529
\bibitem[Mastrodemos, \& Morris(1999)]{mas99}
  Mastrodemos, N., \& Morris, M.\ 1999, \apj, 523, 357
\bibitem[Maxted et al.(2009)]{max09}
 Maxted, P.~F.~L., G\"ansicke, B.~T., Burleigh, M.~R., Southworth,
 J., Marsh, T.~R., Napiwotzki, R., Nelemans, G., \& Wood, P.~L.\
 2009, \mnras, 400, 2012
\bibitem[Mayer et al.(2007)]{may07} Mayer, L., Kazantzidis, S., Madau, P., Colpi, M., Quinn, T., \& Wadsley, J.\ 2007, Science, 316, 1874
\bibitem[Menou et al.(2001)]{men01}
 Menou, K., Haiman, Z., \& Narayanan, V.\ K.\ 2001, \apj, 558, 535
\bibitem[Milosavljevi{\'c} \& Merritt(2003)]{mil03}
 Milosavljevi{\'c}, M., \& Merritt, D.\ 2003, \apj, 596, 860
\bibitem[Narayan(2000)]{nar00} Narayan, R.\ 2000, \apj, 536, 663
\bibitem[Namouni(2010)]{nam10} Namouni, F.\ 2010, \mnras, 401, 319
\bibitem[Nordhaus \& Blackman(2006)]{nor06}
 Nordhaus, J., \& Blackman, E.~G.\ 2006, \mnras, 370, 2004
\bibitem[Ostriker et al.(1992)]{ost92}
   Ostriker, E.\ C., Shu, F.\ H., \& Adams, F.\ C.\ 1992, \apj, 399, 192
\bibitem[Ostriker(1999)]{ost99} Ostriker, E.~C.\ 1999, \apj, 513, 252
\bibitem[Rephaeli \& Salpeter(1980)]{rep80} Rephaeli, Y., \& Salpeter, E.~E.\ 1980, \apj, 240, 20
\bibitem[Ricker \& Taam(2008)]{ric08}
 Ricker, P.~M., \& Taam, R.~E.\ 2008, \apj, 672, L41
\bibitem[Ruderman \& Spiegel(1971)]{rud71} Ruderman, M.~A., \& Spiegel, E.~A.\ 1971, \apj, 165, 1
\bibitem[Ruffert(1993)]{ruf93}
  Ruffert, M.\ 1993, \aap, 280, 141
\bibitem[Sandquist et al.(1998)]{san98}
  Sandquist, E.~L., Taam, R.~E., Chen, X., Bodenheimer, P., \& Burkert,
  A.\ 1998, \apj, 500, 909
\bibitem[S{\'a}nchez-Salcedo \& Brandenburg(2001)]{san01} S{\'a}nchez-Salcedo, F.~J., \& Brandenburg, A.\ 2001, \mnras, 322, 67
\bibitem[Shu(1992)]{shu92} Shu, F.\ H.\ 1992, The Physics of Astrophysics. II.
   Gas Dynamics (Mill Valley: Univ.\ Science Books)
\bibitem[Stone \& Norman(1992)]{sto92}
   Stone, J.\ M., \& Norman, M.\ L.\ 1992, \apjs, 80, 753
\bibitem[Taam \& Sandquist(2000)]{taa00}
 Taam, R.\ E., Sandquist, E.\ L.\ 2000, \araa, 38, 113
\bibitem[Tanaka et al.(2002)]{tan02}
 Tanaka, H., Takeuchi, T., \& Ward, W.~R.\ 2002, \apj, 565, 1257
\bibitem[Villaver \& Livio(2009)]{vil09}
 Villaver, E., \& Livio, M.\ 2009, \apj, 709, L81
\bibitem[Ward(1997)]{war97}
 Ward, W.~R.\ 1997, ICARUS, 126, 261
\end{thebibliography}
\end{document}